\documentclass[useAMS,usenatbib]{mn2e}
\usepackage{epsfig,amsmath,amssymb,amsfonts,mathrsfs,latexsym,graphicx}
\begin{document}

\title[A Post Maximum Supernova Radiation Transport
  Code]{NERO $-$ A Post
  Maximum Supernova Radiation Transport Code}

\author[Maurer et al.]{I. Maurer$^{1,**}$, A. Jerkstrand$^{2,3}$,
  P. A. Mazzali$^{1,4}$, S. Taubenberger$^{1}$, S. Hachinger$^{1}$,
  \and  M. Kromer$^1$, S. Sim$^5$, and W. Hillebrandt$^{1}$\\
\\$^1$ Max Planck Institut f\"ur
  Astrophysik, Karl-Schwarzschild-Str.1, 85741 Garching, Germany
\\$^{2}$ Department of Astronomy, Stockholm University, Alba Nova
University Centre, SE-106 91 Stockholm
\\$^3$ The Oskar Klein Centre, Stockholm University
\\$^4$ National Institute for Astrophysics-OAPd, Vicolo
dell'Osservatorio, 5, 35122 Padova, Italy
\\$^{5}$ Research School of Astronomy and Astrophysics, Mount Stromlo
Observatory, Cotter Road, Weston Creek, ACT 2611, Australia
\\
\\$^{**} maurer@mpa$-$garching.mpg.de$
}

\maketitle

\begin{abstract}
The interpretation of supernova (SN) spectra is essential for deriving
SN ejecta properties such as density and composition, which in turn
can tell us about their
progenitors and the explosion mechanism. A very large number of atomic
processes are important for
spectrum formation. Several tools for calculating SN spectra exist, but they
mainly focus on the very early or late epochs. The intermediate phase,
which requires a NLTE treatment of radiation transport has rarely been
studied.

In this paper we present a new SN radiation transport code, {\sc nero},
which can look at those epochs.
All the atomic processes are treated in full NLTE, under a steady-state
assumption. This is a valid approach
between roughly 50 and 500 days after the explosion depending on SN
type. This covers the
post-maximum photospheric and the early and the intermediate nebular
phase.

As a test, we compare {\sc nero} to the radiation transport code of
\citet{Jerkstrand11} and to the nebular code of \citet{Mazzali01}. All three codes
have been developed independently and a comparison provides a 
valuable opportunity to investigate their reliability.
Currently, {\sc nero} is one-dimensional and can be used for predicting spectra of
synthetic explosion models or for deriving SN properties by spectral
modelling. To demonstrate this, we study the spectra of the
'normal' SN Ia 2005cf between 50 and 350 days after the explosion and
identify most of the common SN Ia line features at post maximum epochs.
\end{abstract}


\maketitle

\section{Introduction}
\label{intro}
Supernovae are classified by their spectra. The most common types are SNe
Ia, Ib, Ic, IIb and II \citep[e.g.][]{Turatto07}. This classification links directly to
the explosion mechanism, which can be either a thermonuclear
explosion or the collapse of a massive star. The connection can be
established since the progenitor and the explosion mechanism leave a
unique imprint on the resulting SNe, which is reflected by their
spectra.

In turn, one can learn about the properties of
observed SNe by spectral analysis. The
earliest attempts to interpret the spectra of SNe reach back to the
very beginning of SN astronomy, when SNe I were classified as
hydrogen poor and SNe II as hydrogen rich \citep[e.g.][]{Minkowski41}. Since then, the
interpretation of SN spectra has been steadily refined and several
advanced tools to study SN spectra formation exist today (see below).

SN spectra are usually grouped into photospheric and nebular. 
It is not trivial to distinguish between the two phases on
physical grounds since most radiation process important during the
earliest phases play also a role at late times. However, nebular phase
spectra are dominated by clear forbidden-line emission features, which
become dominant at about 200 days after the explosion. Although
photo-excitation and -ionisation processes are important at later
epochs \citep[e.g.][also see this paper]{Li96,deKool98,Jerkstrand11}, they are
absolutely dominant at earlier epochs.

Directly after the explosion the gas is
in local thermodynamic equilibrium (LTE), which simplifies
radiation transport. However, shortly after maximum light at latest
non-local thermodynamic equilibrium (NLTE) effects become important. Very little
attention has been paid to these intermediate epochs so far, which require a
full NLTE treatment of radiation transport. 

Codes treating the early phase have for example been developed by
\citet[][]{Mazzali93,Lucy99,Kasen06,Kromer09}. These early time codes rely at
least partially on the LTE
assumption and do not treat forbidden-line emission. For later epochs
numerical treatments have for example been developed by
\citet[][]{Axelrod80,Ruiz92,Kozma92,Kozma98,Eastman93,deKool98,Mazzali01,Jerkstrand11}. In
addition to those spectral codes specified for SNe, there exist others
\citep[e.g.][]{Pauldrach96,Hauschildt97,Hillier98} which can be used for calculating
SN spectra. The list presented here is a selection and is far from complete. 

Here we present a new non-thermal equilibrium radiation transport ({\sc nero}) code
which addresses especially the intermediate epochs at
about 50 $-$ 200 days after the explosion. We can also treat later epochs, as long as the gas
shows no strong deviations from steady-state (see Section \ref{code}).

In Section \ref{code} we describe the new code. In Section
\ref{results} we compare {\sc nero} to the
steady-state radiation transport code of \citet{Jerkstrand11} and to the
nebular code of \citet{Mazzali01}. We also compare synthetic spectra
obtained by using {\sc nero} on a SN Ia W7 model \citep{Nomoto84} with observations of the proto-typical
'normal' SN Ia 2005cf at epochs between 50 and 350 days after the
explosion. In Section \ref{disc} our results are discussed.  

\section{Code description}
\label{code}  
{\sc nero} is built on
the assumption that the SN ejecta are in a steady-state, 
i.e. heating and ionisation are balanced by
cooling and recombination, which is
valid between roughly 50 $-$ 500 days after the explosion
\citep[e.g.][]{Axelrod80}. These limits however depend strongly on the ejecta density and
composition. Under this assumption the spectral emission can be
calculated at any epoch without considering the radiation history of
the gas. The phase we can treat begins when the
SN light-curve starts to follow the decay of Ni and Co, which means
that the gas is in thermal equilibrium and ends when
the gas falls out of ionisation equilibrium or when adiabatic cooling
becomes important \citep[e.g.][]{Kozma92}.\\
\begin{figure} 
\begin{center}
\includegraphics[width=6.5cm, clip]{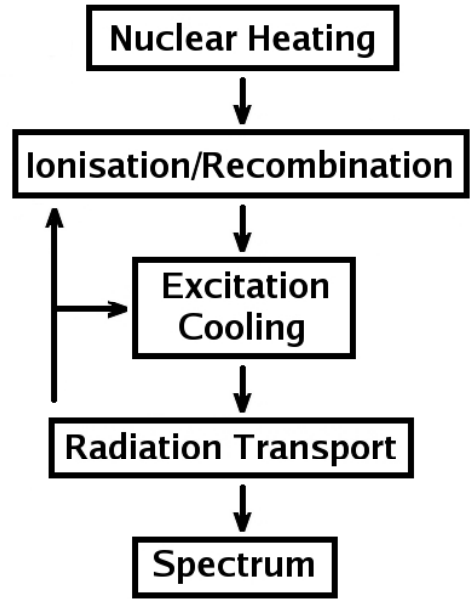}
\end{center}
\caption{Code scheme of {\sc nero}.}
\label{figcode}
\end{figure}
In a first step, the energy deposition from the decay of
radioactives is calculated, from which we derive the non-thermal
electron ionisation and excitation rates. Using an initial guess for
the ionisation and excitation state and the temperature of the gas we derive its new ionisation and
excitation state by solving the statistical equilibrium equations. From the ion fractions and occupation numbers one
can calculate the radiation field. This radiation field is
transported through the SN ejecta using a ray-tracing transport
scheme, from which we derive photo-ionisation and excitation
rates. The ionisation and excitation states are iterated until
the electron density and temperature and the radiation field have
converged. A scheme of {\sc nero} is shown in Figure \ref{figcode}.

Currently, {\sc nero} is one-dimensional, but a three-dimensional
version may be available in the future. For a one-dimensional
model consisting of about 20 radial shells, one calculation takes
a few minutes on a standard desktop computer, depending on
the composition and density of the model and the desired
resolution. All synthetic spectra obtained for this paper using
{\sc nero} required a calculation time of the order of 5 minutes,
which is excellent for a fully-fledged NLTE radiation
transport code. Still, we
plan to parallelise and further optimise {\sc nero} in the future, to
make  modelling of observed SN spectra as efficient as possible. 

\subsection{Radioactive deposition}
\label{heating}
{\sc nero} can treat radioactive deposition by $^{56,57}$Co and by
$^{56,57}$Ni. The half-lives of these isotopes can for example be
found in \citet{Seitenzahl09}. Including other radioactives would be
simple. The energy deposition process is treated with a
Monte Carlo approach, where Compton scattering,
pair-creation and photo-ionisation are taken into account, as for example
described in \citet{Sim08}. We calculate the spectrum of the emerging
$\gamma$-radiation. However, there are no observations available to
compare with. The accuracy of our deposition routine
has been verified
by comparing the energy deposition rates to calculations performed with the spectral codes of
\citet{Mazzali01} and \citet{Jerkstrand11}. From the deposited luminosity we calculate
non-thermal ionisation and excitation rates as described in
\citet{Maurer10b} and \citet{Maurer10c}. To evaluate the non-thermal ionisation rates
we estimate the fraction of the deposited luminosity causing
ionisation using the Bethe approximation \citep[e.g.][]{Axelrod80}. The non-thermal
excitation rates are calculated using the optical approximation
\citep[e.g.][]{Axelrod80,Rozsnyai80}. It was found by \citet{Maurer10c} that this treatment of
non-thermal ionisation and excitation is accurate to at least 20\% for
hydrogen and helium. We also compared our non-thermal electron ionisation and excitation rates
for various ions to rates calculated with the radiation transport code of \citet{Jerkstrand11}, which
makes use of the Spencer-Fano approach of \citet{Kozma92}. We find
overall good agreement, with deviations usually of the order of 10\% and
$\sim$ 50\% in the worst case. Especially for the iron-group
elements the non-thermal rates are
strongly affected by inaccurate or poorly known
atomic data and a high degree of uncertainty has to be
accepted. This may be the largest source of uncertainty in all
spectral calculations of SNe Ia after maximum light.

\subsection{Ionisation \& Recombination}
\label{ion}
The ionisation state of the gas is calculated under the assumption of statistical
equilibrium, balancing non-thermal electron ionisation,
photo-ionisation and radiative and di-electronic recombination with
the charge-exchange reactions listed in \citet{Swartz94}. The
non-thermal electron rates are obtained as described in Section
\ref{heating}. The photo-ionisation rates are obtained from ray-tracing
radiation transport (described below) using the ground state
photo-ionisation cross-sections from TIPTOP
Base\footnote{http://cdsweb.u-strasbg.fr/OP.htx} and a simple approximation
for the lowest 40 excited states. Above the corresponding ionisation
thresholds, the excited-state photo-ionisation cross-sections
are assumed to be constant fractions of the
ground-state cross-sections. These fractions decrease with increasing main
quantum number of the excited states. All rates are corrected for stimulated
recombination. We do not treat photo-ionisation from states higher
than 40 since they seem to have no noticeable influence on the
spectra at the epochs of interest (50 $-$ 500 days). We use the radiative and
di-electronic total recombination rates of \citet{Mazzotta98}. Ground state
recombination rates are taken from \citet{Aldrovandi73} or are set to 10\% of the
total radiative recombination rate, if not available. All the electrons
recombining into excited states are distributed equally to the lowest 40
excited states from where they undergo the complete NLTE process. We plan to improve our excited
state ionisation and recombination data base in the future.

\subsection{Excitation}
\label{excitation}
We use the line data collection of \citet{Kurucz95}, which roughly
contains 25.000 atomic levels and 500.000 lines. In principle, we can treat all elements
from H to Ni and all ions from ionisation state I $-$ III (IV is taken into account for
the ionisation equilibrium but does not contribute to the radiation
field). However, the atomic data are poor for many ions. The atomic excitation states
are calculated by solving a rate matrix \citep[e.g.][]{Axelrod80},
including non-thermal electron excitation rates (see Section
\ref{heating}), photo-ionisation and excitation rates obtained from ray-tracing
radiation transport (see below) using the Sobolev approximation,
spontaneous radiative de-excitation, recombination into excited states,
thermal electron (de-) excitation,
continuum destruction \citep[e.g.][]{Chugai87,Li95} and two-photon emission (TPE) of H~{\sc i}(2s$^1$S)
and He~{\sc i}(2s$^{1,3}$S) [e.g. see
\citet[][]{Kaplan72,Drake69} for TPE rates].

With the electron density and temperature obtained from the iteration
process, we calculate collisional (de-) excitation rates. Collisional
data are taken from TIPTOP and CHIANTI
databases\footnote{http://www.chiantidatabase.org/} but also from
other sources
\citep[e.g.][]{Berrington82,Hayes84,Berrington88,Mauas88,Scholz90,Callaway94,Melendez07,Bautista09}.
Since our collisional database is far from complete we plan to
regularly add and update collisional atomic data. If not available,
the collision strengths are approximated
\citep[e.g.][]{VanRegemorter62,Axelrod80}. However, for several
hundreds of lines there is collisional data 
from the literature. A serious problem at intermediate epochs, especially for treating SNe Ia, is
the absence of reliable collisional data for Co. 

\subsection{Radiation Transport}
From the ionisation and excitation states obtained in the previous steps we calculate a
radiation field, which is represented by a
certain amount of photon packets (typically, a few 100.000
per shell in total). These are sent through the SN
envelope in random directions. On their way out they propagate on
straight lines and encounter bound-bound and bound-free absorption and
electron-scattering. The probabilities for line scattering are
calculated in the Sobolev approximation. While transported, the
photon packets lose parts of their energy according to the
respective optical depths (or change their direction after electron
scattering), and can be absorbed to 100\%
if the optical depth is much larger than one. The absorbed
photon packets are
re-emitted in random directions in the next iteration step after taking part in the
NLTE excitation matrix calculation (Section \ref{excitation}). Therefore, all the absorbed energy
undergoes the full NLTE process, including fluorescence, up- and down-ward electron
collisions, photo-ionisation, recombination, continuum destruction and
two-photon emission.

\section{Code results}
\label{results}

\subsection{Comparison to other codes}
\label{cc}

\begin{table}
 \centering
  \caption{SN 1994I. (A) RTM one-zone model (B) {\sc nero}
    one-zone model (C) {\sc nero} small-scale separation model}
  \begin{tabular}{ccccccccc}
  \hline
   & C & O & Na & Mg & S & Ca & Ni \\
  & M$_\odot$ & M$_\odot$  & M$_\odot$ &  M$_\odot$ & M$_\odot$ &
  M$_\odot$ &  M$_\odot$  \\
 \hline
  A  & 0.09 & 0.2 & 0.0002 & 0.002 & 0.02 &  0.001 & 0.07  \\
  B  & 0.2 & 0.6 & 0.01 & 0.1 & 0.03 & 0.0004 & 0.03 \\
  C  & 0.07 & 0.2 & 0.05 & 0.1 & 0.01 & 0.002 & 0.05 \\
\hline
\end{tabular}
\label{tab1}
\end{table}

In this section we compare {\sc nero} to the radiation transport code of
\citet{Jerkstrand11} [RTJ] and to the nebular code of
\citet{Mazzali01} [RTM]. RTJ is a steady-state radiation
transport code. Although developed independently, it is built on very similar physical
assumptions as {\sc nero}. While {\sc nero} treats the NLTE process completely, RTJ does not allow up-ward
excitation of photo-excited levels. However, at least at the
late epochs, which have been chosen for the code comparison, this seems to have no
observable influence on the synthetic spectra (see below).  RTJ has been applied to study the
very late phase of SN 1987A so far \citep{Kjaer10,Jerkstrand11}. 

RTM is a nebular code based on the ideas of
\citet{Axelrod80} and \citet{Ruiz92}, therefore neglecting all radiation transport effects. It has widely been
used in the literature \citep[e.g.][]{Silverman09,Mazzali10} to study nebular spectra of all
types of SNe. 

While {\sc nero} and RTJ are currently available in one-dimensional
versions only, there are three-dimensional versions of RTM
\citep{Maeda02,Maurer10a}.

A code comparison is interesting, since all three codes have been
developed independently and
use different numerical methods, physical assumptions and partially different atomic data. 

\begin{figure} 
\begin{center}
\includegraphics[width=8.5cm, clip]{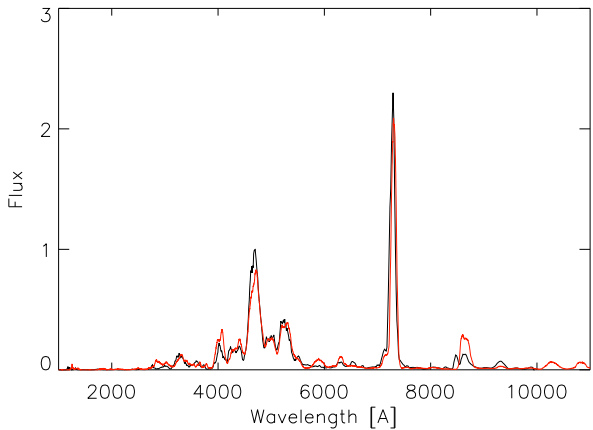}
\end{center}
\caption{Synthetic spectra of the \citet{Woosley94} 13C model at 200
  days after the explosion. The
  RTJ calculation is shown in black, while the red
  curve was produced using {\sc nero}. The agreement is excellent.}
\label{comp13c200}
\end{figure}

\begin{figure} 
\begin{center}
\includegraphics[width=8.5cm, clip]{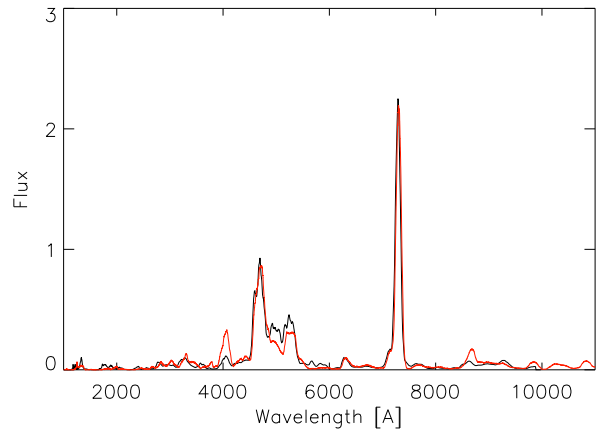}
\end{center}
\caption{Synthetic spectra of the \citet{Woosley94} 13C model at 400
  days after the explosion. The
  RTJ calculation is shown in black, while the red
  curve was produced using {\sc nero}. There is some
  disagreement, especially in the Fe dominated region around 5000 \AA\
  but in general the agreement is excellent.}
\label{comp13c400}
\end{figure}

For the comparison of RTJ and {\sc nero}, we chose the 13C model of \citet{Woosley94}.
At 200 and 400 days after the
explosion excellent agreement (see
Figures \ref{comp13c200} \& \ref{comp13c400}) is observed. The 13C model is found
to show very strong Ca {\sc ii} emission in both calculations, which can
however be explained to be
a mixing effect. In the 13C model all the Ca
is mixed with the other elements microscopically. Since Ca {\sc ii}
has a low excitation potential and large collision strengths it
radiates strongly, if it is mixed into large amounts of hydrogen, helium or oxygen.

\begin{figure} 
\begin{center}
\includegraphics[width=8.5cm, clip]{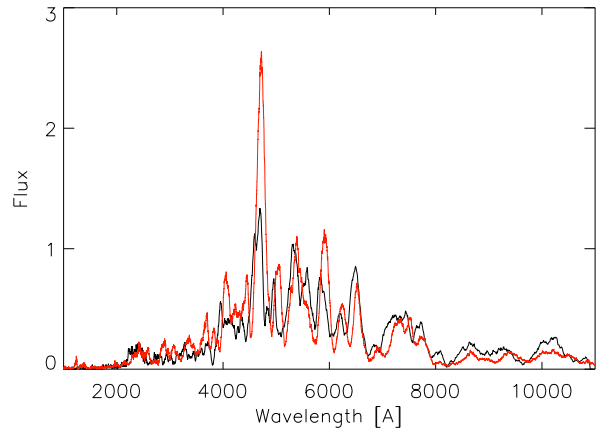}
\end{center}
\caption{Synthetic spectra of the W7 model at 94
  days after the explosion. The
  RTJ calculation is shown in black, while the red
  curve was produced using {\sc nero}. The agreement is reasonable, apart
  from a Fe {\sc iii} feature around 4600 \AA\ and a Co {\sc iii}
  feature around 5900 \AA .}
\label{compw794}
\end{figure}

\begin{figure} 
\begin{center}
\includegraphics[width=8.5cm, clip]{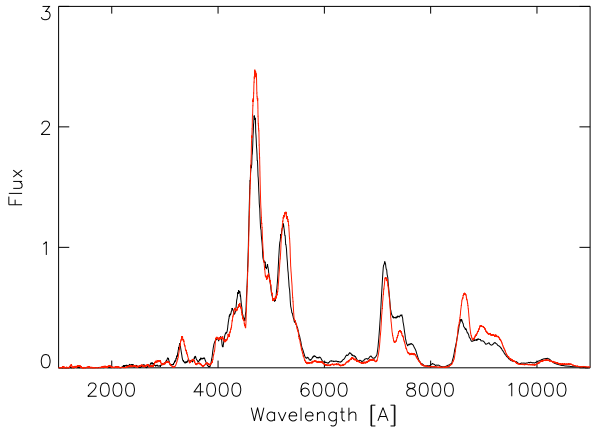}
\end{center}
\caption{Synthetic spectra of the W7 model at 338
  days after the explosion.  The
  RTJ calculation is shown in black, while the red
  curve was produced using {\sc nero}. There is some
  disagreement, but in general the agreement is good.}
\label{compw7338}
\end{figure}

Another comparison of {\sc nero} and RTJ was performed for a Type Ia
W7 model \citep{Nomoto84} at 94 and 338 days after the explosion. The
agreement at both epochs is good (see Figures \ref{compw794} \&
\ref{compw7338}). However, at 94 days after the explosion there is a strong
deviation around 4600 \AA\ and 5900 \AA . These features are caused by
Fe {\sc iii} and Co {\sc iii}, respectively. The fraction of these
ions is similar in both calculations, which means that the ion abundance
cannot be a main reason for this differences. Since both features are
dominated by collisional excitation from ground levels and since the
electron temperature and the density are similar in both calculations, it is likely that the differing
sets of atomic data used in {\sc nero} and RTJ are responsible for most
of the deviation. Also, there is is an important
fraction ($\sim$ 15\% at 5000 km/s, increasing with velocity) of Fe {\sc iv} and
Co {\sc iv} in the {\sc nero} calculation, which is neglected by
RTJ at the moment. The recombination of these ions (and
photo-ionisation of Fe {\sc iii} and Co {\sc iii}) influences
the radiation field and the cascading of UV radiation can
hardly be followed in detail. It is interesting to note that the Fe
{\sc iii} feature produced with RTJ is more consistent with observed SNe
Ia spectra, while this is true for the Co {\sc iii} feature produced
with {\sc nero} (see below).

\begin{figure} 
\begin{center}
\includegraphics[width=8.5cm, clip]{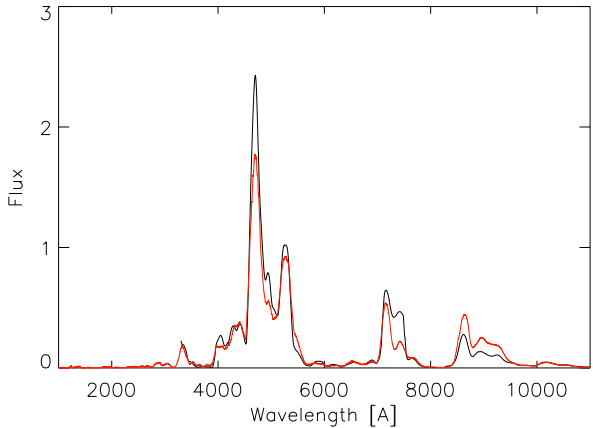}
\end{center}
\caption{Synthetic spectrum of the W7 model at 338
  days after the explosion. The RTM
  calculation is shown in black, while the red
  curve was produced using {\sc nero}. In general the agreement is
  good.}
\label{compw7}
\end{figure}

For the comparison of RTM and {\sc nero} we chose a Type Ia W7 model \citep{Nomoto84}
and the 'standard' Type Ic SN 1994I \citep{Sauer06}.
The comparison of spectra obtained with  RTM and {\sc nero} for the W7 model \citep{Nomoto84} 
shows reasonable  agreement (see Figure \ref{compw7}). The core of the W7 model, which is
observed in the nebular phase, consists of almost pure Fe (from
$^{56}$Ni decay). In {\sc nero} Fe {\sc i} is
photo-ionised almost completely. The ratio of Fe {\sc ii} to Fe {\sc iii} is dominated
by non-thermal electron ionisation at late epochs, which is also treated by RTM. In RTM the fraction of Fe {\sc i}
is always set to zero. Therefore, the ionisation state (and hence
the electron density) of a pure Fe
core obtained by {\sc nero} and RTM is similar for this model. The
electron temperature of the Fe plasma obtained by {\sc nero} is about
10\% lower than that
obtained with RTM, because of the presence of
excitation processes (photo-excitation, excitation by excited-state
recombination, non-thermal electron excitation, more collisional
transitions) which are neglected in RTM calculations.

\begin{figure} 
\begin{center}
\includegraphics[width=8.5cm, clip]{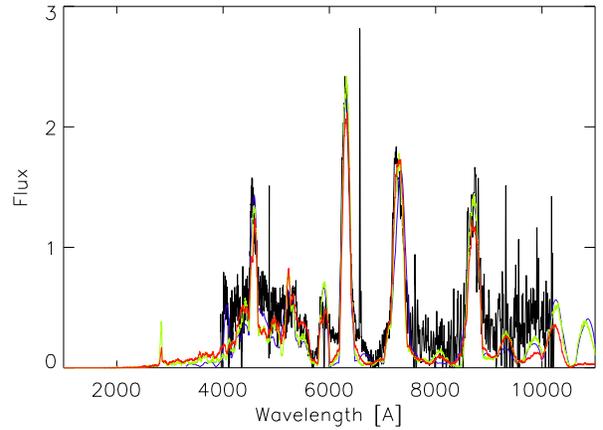}
\end{center}
\caption{The spectrum of SN 1994I at about 159 days after the
  explosion is shown in black. The spectrum is taken from
  \citet{Sauer06} and an identification of the most important
  emission lines can be found there. The RTM model is shown in blue, while the {\sc
    nero} models 'B' and 'C' are shown in green and red,
  respectively. All spectra have been scaled by $2.2 \cdot
  10^{15}$ cm$^{2}$ s ergs$^{-1}$. All models seem to fit the observed spectrum equally well. Their
  properties are listed in Table \ref{tab1}.}
\label{comp94I}
\end{figure}

Since RTM has exclusively been used to derive SN core ejecta properties (and
not for predicting spectra from explosion models), 
we also compare modelling results of RTM and
{\sc nero}. For this comparison, we chose SN Ic 1994I, which has been
studied in detail by \citet{Sauer06}. They derived a total
core mass ($v <$ 5500 km/s) of 0.43 M$_\odot$ and a $^{56}$Ni mass of
0.07 M$_\odot$, which was also found to be consistent with the observed
light curve of SN 1994I. We fit the observed spectrum with one-zone
models (see Figure \ref{comp94I}), as it was done by \citet{Sauer06} [see Table
\ref{tab1}, model 'A' (RTM) \& 'B' ({\sc nero})]. 

In general, the masses estimated for Na, Mg, Si, S and
Ca deviate in RTM and {\sc nero} calculations.  This is
expected, since these elements are strongly influenced by
photo-ionisation, which is not treated in RTM. Therefore, one derives more mass for the elements
which are estimated from the neutral component (e.g. Na, Mg) and less
for those which are estimated from ionised states (e.g. Ca) using {\sc
  nero}.

Another important difference is found for the estimate of the total
and the $^{56}$Ni mass. These two quantities can be considered as the
main properties of any SN and should be in unison with its
light curve. While the RTM calculation is consistent with the
light curve \citep[see][]{Sauer06}, the {\sc nero} calculation seems
to show too much total mass and too little $^{56}$Ni (see Table
\ref{tab1}, model 'B') to be consistent
with the light curve modelling results of \citet{Sauer06}.

This can have several reasons. Unfortunately, the intrinsic uncertainty of
light curve modelling is hard to estimate and there is always some
degeneracy between $^{56}$Ni and total mass, especially when radiation
transport is treated in rough approximation. 

However, assuming that the $^{56}$Ni and total mass derived by \citet{Sauer06}
are correct, the discrepancy of the main properties can be explained by the over-simplified
input model that was used in the calculation. It is well
known, that mixing of the ejecta does have an important effect on the
resulting spectra (also see above). 

While for the code comparison of RTM and {\sc nero} a
one-zone model has been used, a real SN is certainly more
complex. Apart from large scale asymmetries (which are not
necessarily expected in
SN 1994I), the ejecta can be structured on much smaller scales. If $^{56}$Ni
and other elements are separated, the ratio of Fe and other element lines
changes. This, in turn, influences the estimate of the ejecta properties.

Mixing can influence the spectrum in two ways. First, separation of $^{56}$Ni and other
elements reduces the $\gamma$ heating in the non-radioactive
zones. However, in the intermediate nebular phase this effect is weak,
since the $\gamma$ opacity is low and positrons do not dominate
yet. More importantly, when separated, carbon or oxygen rich zones
cannot cool via Fe emission lines. This means, that separating $^{56}$Ni from other
elements in the nebular phase can lead to stronger emission of those elements than with
perfect mixing, which may seem counter-intuitive on a first glance.

To demonstrate this, we model SN 1994I using {\sc nero} again, this
time separating Fe and O in several thin shells to simulate a
separation of the ejecta on small scales. The derived $^{56}$Ni and total mass
changes strongly (see Table \ref{tab1}, model 'C') and becomes more
consistent with the light curve estimate of \citet{Sauer06}. Of
course, such an approach is highly degenerate and a broad variety of
modelling results is possible.
 
This causes an unfortunate situation. On the one hand, detailed
knowledge of the mixing of the ejecta on large and on small scales is necessary
to derive the main ejecta
properties. On the other hand, this information is poorly constrained
from observations and explosion models, especially on the small
scales. Therefore, mixing poses a
problem for deriving ejecta properties of stripped-envelope
core-collapse SNe from modelling. Within these
uncertainties, RTM seems appropriate to derive the main properties of
SNe. For {\sc nero} calculations, more elaborate input models seem
to be necessary to become consistent with the light curve modelling.

\subsection{Comparison to observations}
\begin{figure} 
\begin{center}
\includegraphics[width=8.5cm, clip]{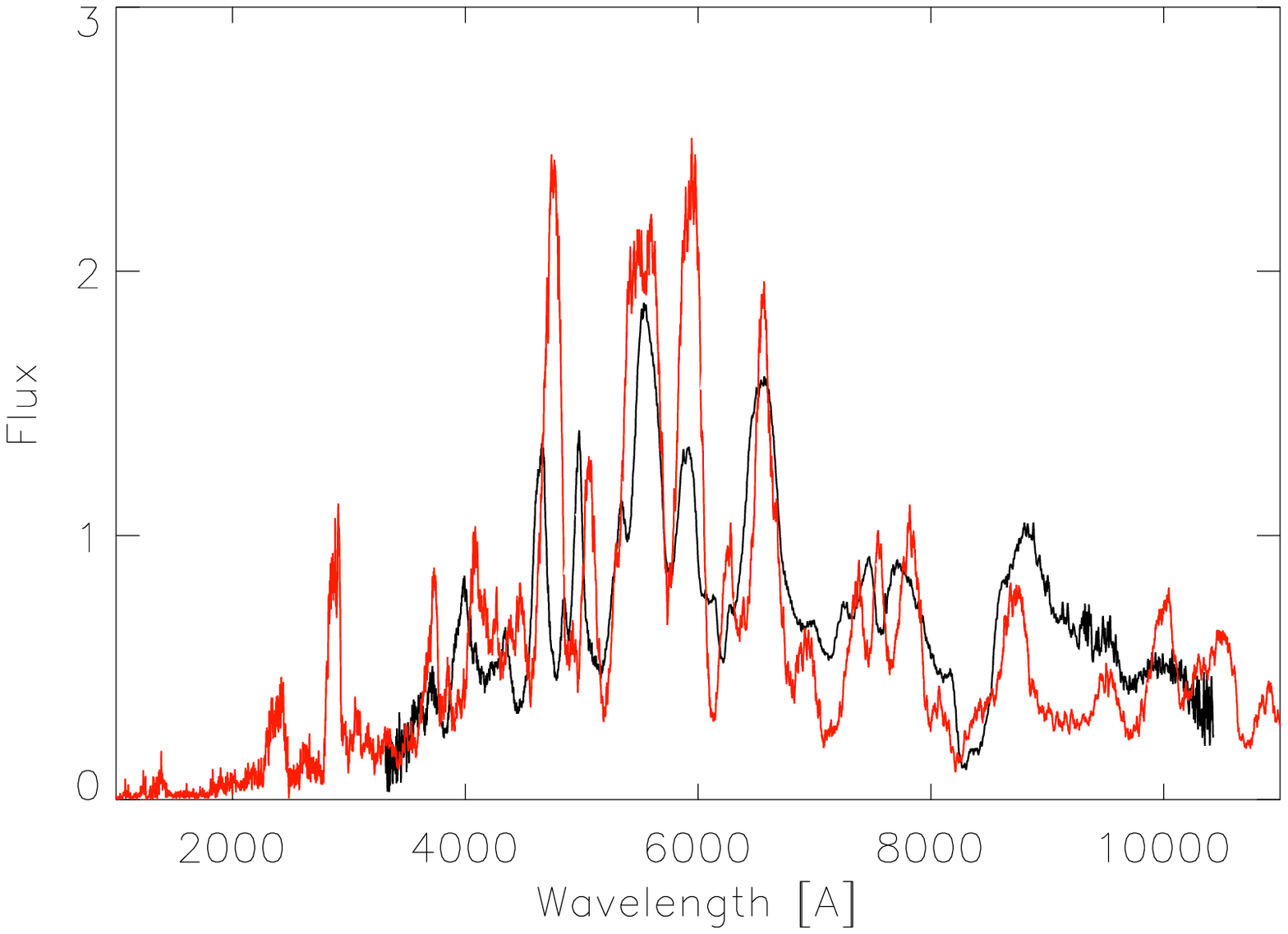}
\includegraphics[width=8.5cm, clip]{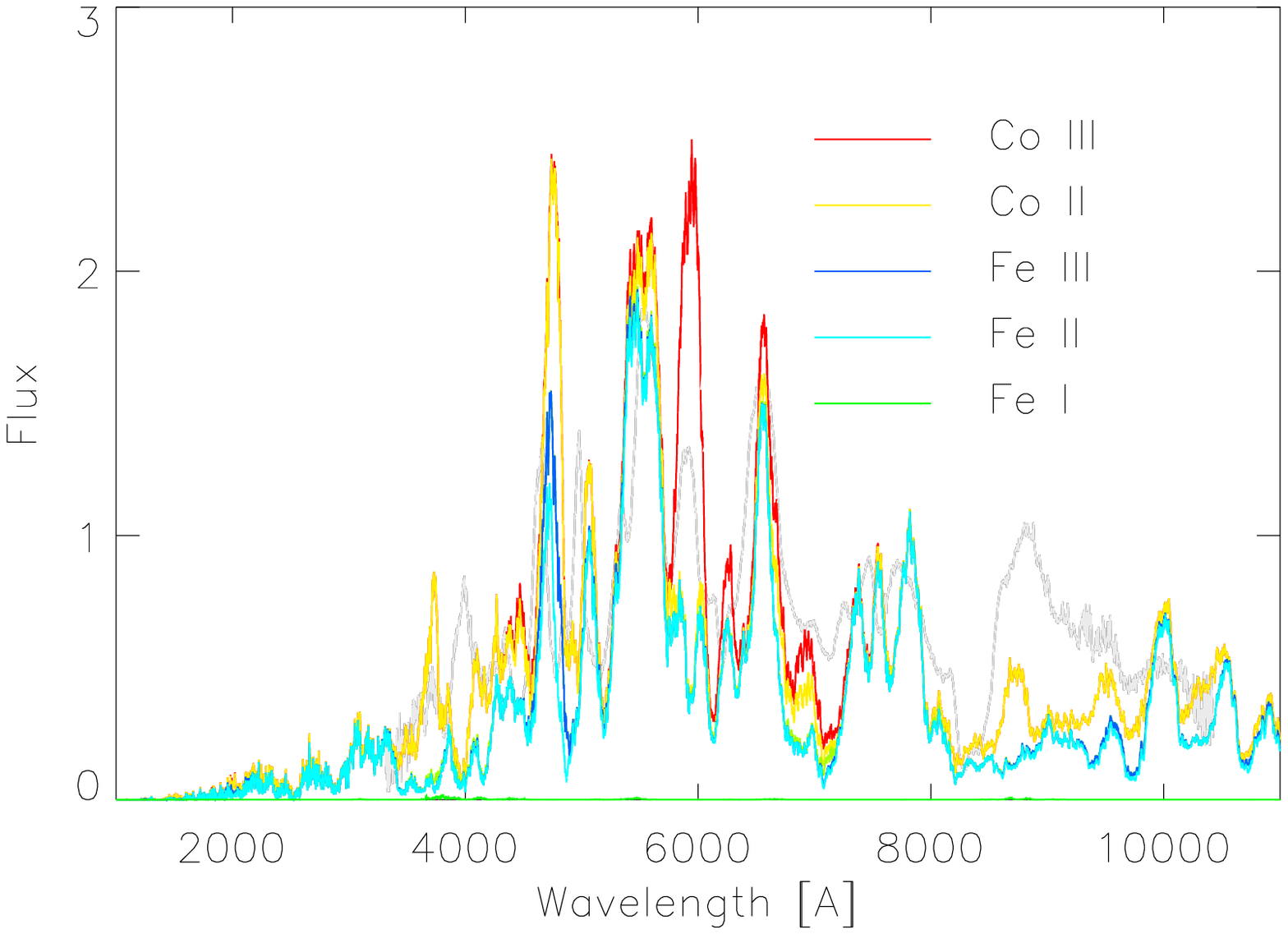}
\end{center}
\caption{The spectrum of SN 2005cf at about 47 days after the
  explosion \citep{Wang09} is shown in black (upper panel) or light
  grey (lower panel). The spectrum was scaled
  by a constant. The coloured curves were
  produced using {\sc nero} on W7. Upper panel: the red line shows the
  total synthetic flux. In general the agreement is good. It is important
to note that the early-time ejecta are dominated by Co {\sc ii} \&
{\sc iii}, which have
poorly known collisional data. Lower panel: the flux of Co {\sc iii}
  (red), Co {\sc ii} (orange), Fe {\sc iii} (dark blue),
  Fe {\sc ii} (light blue) and Fe {\sc i} (green) is shown
  separately. Most features are reproduced well. 
Disagreement is found around 4600 \AA\
(Fe {\sc ii}, Co {\sc ii}), $\sim$ 5900 \AA\ (Co {\sc iii}) and around
$\sim$ 8700 \AA , which is possibly Ca {\sc ii} and [Co {\sc ii}] in the observed spectrum. Our
synthetic spectra contain contributions from elements other than
Fe or Co, which are not shown in the lower
plot.}
\label{05cf47}
\end{figure}

\begin{figure} 
\begin{center}
\includegraphics[width=8.5cm, clip]{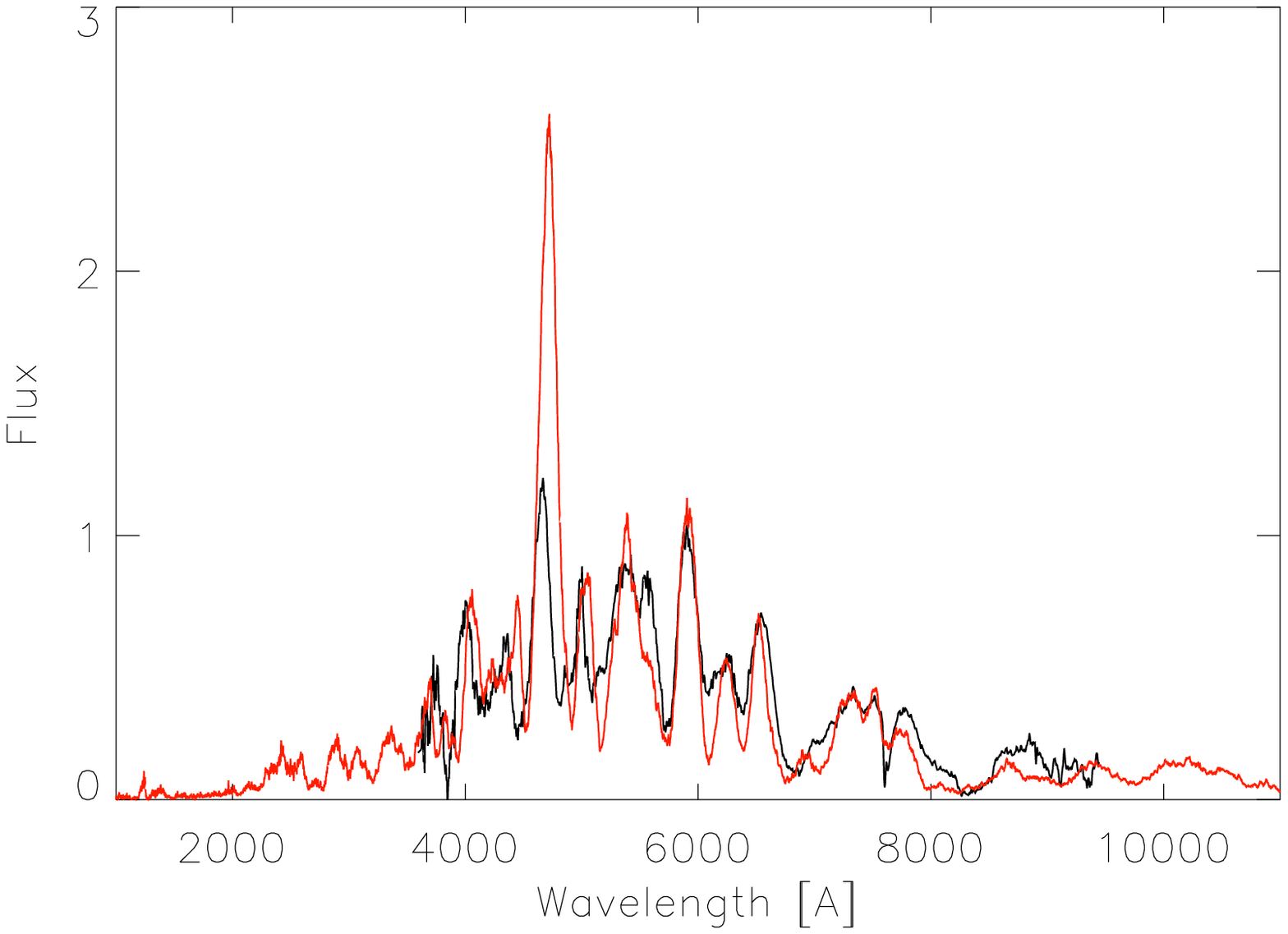}
\includegraphics[width=8.5cm, clip]{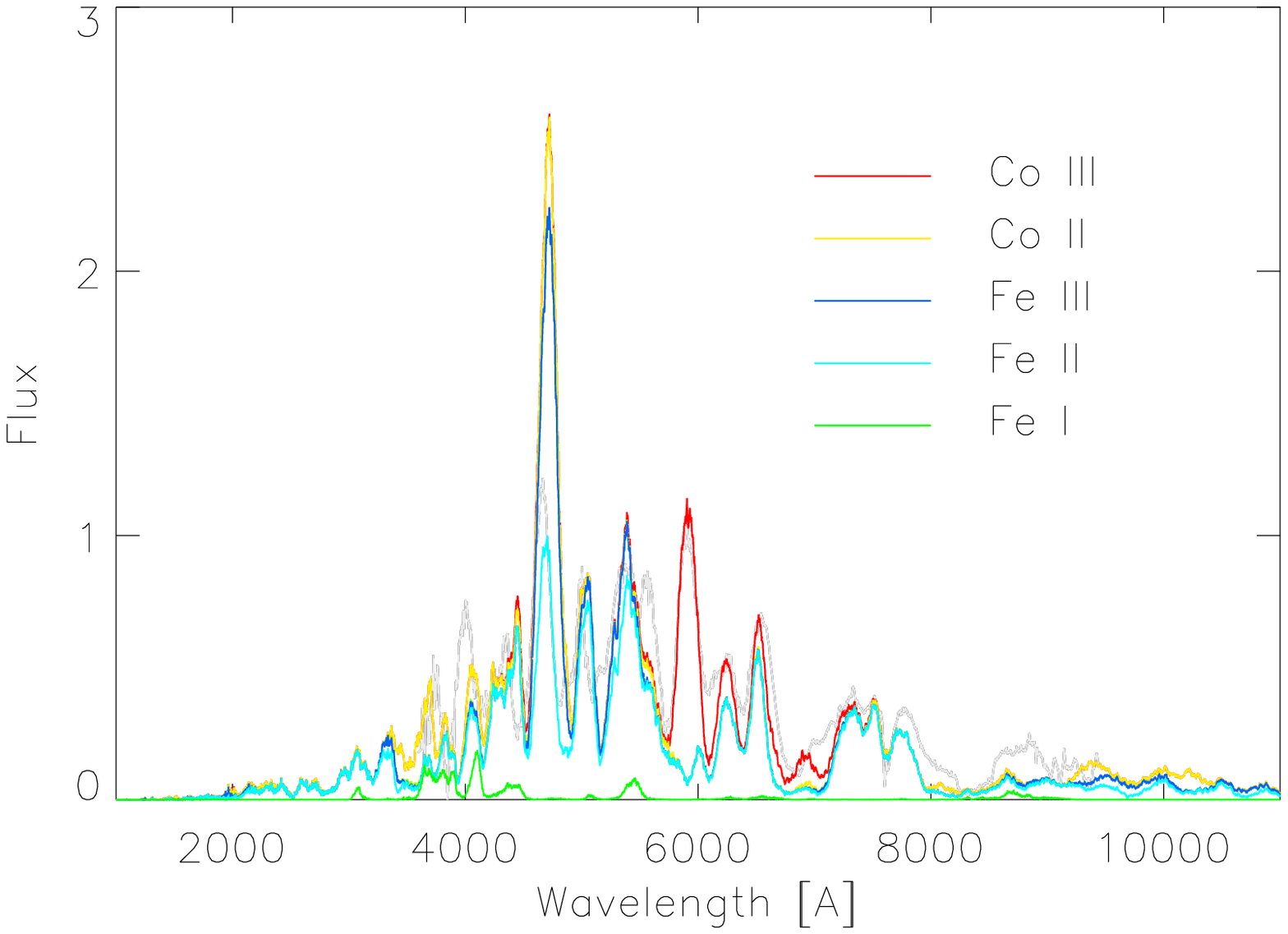}
\end{center}
\caption{The spectrum of SN 2005cf at about 94 days after the
  explosion \citep{Garavini07} is shown in black (upper panel) or light
  grey (lower panel). The spectrum was
  scaled by a constant. The coloured curves were
  produced using {\sc nero} on W7. Upper panel: the red line shows the
  total synthetic flux. In general the
  agreement is excellent apart from the region around 4600 \AA . Lower panel: the flux of Co {\sc iii}
  (red), Co {\sc ii} (orange), Fe {\sc iii} (dark blue),
  Fe {\sc ii} (light blue) and Fe {\sc i} (green) is shown separately. The synthetic
  flux exceeds the observed one at $\sim$ 4600 \AA\ by a factor of
  $\sim$ 3 and is dominated by [Fe {\sc iii}] emission. At the same
  time, the [Co {\sc iii}] feature at $\sim$ 5900 \AA\ matches
  the observed spectrum well.}
\label{05cf94}
\end{figure}

\begin{figure} 
\begin{center}
\includegraphics[width=8.5cm, clip]{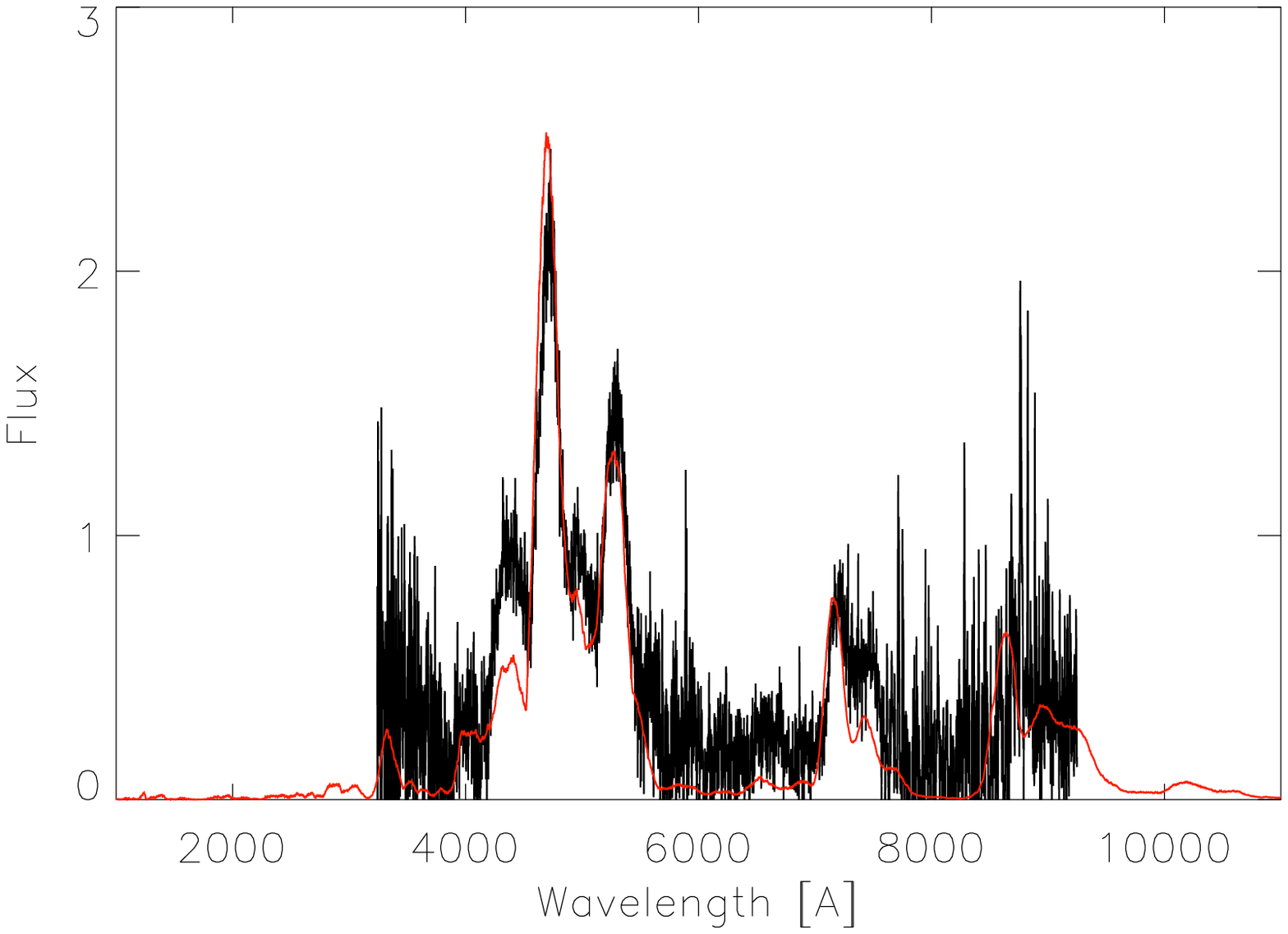}
\includegraphics[width=8.5cm, clip]{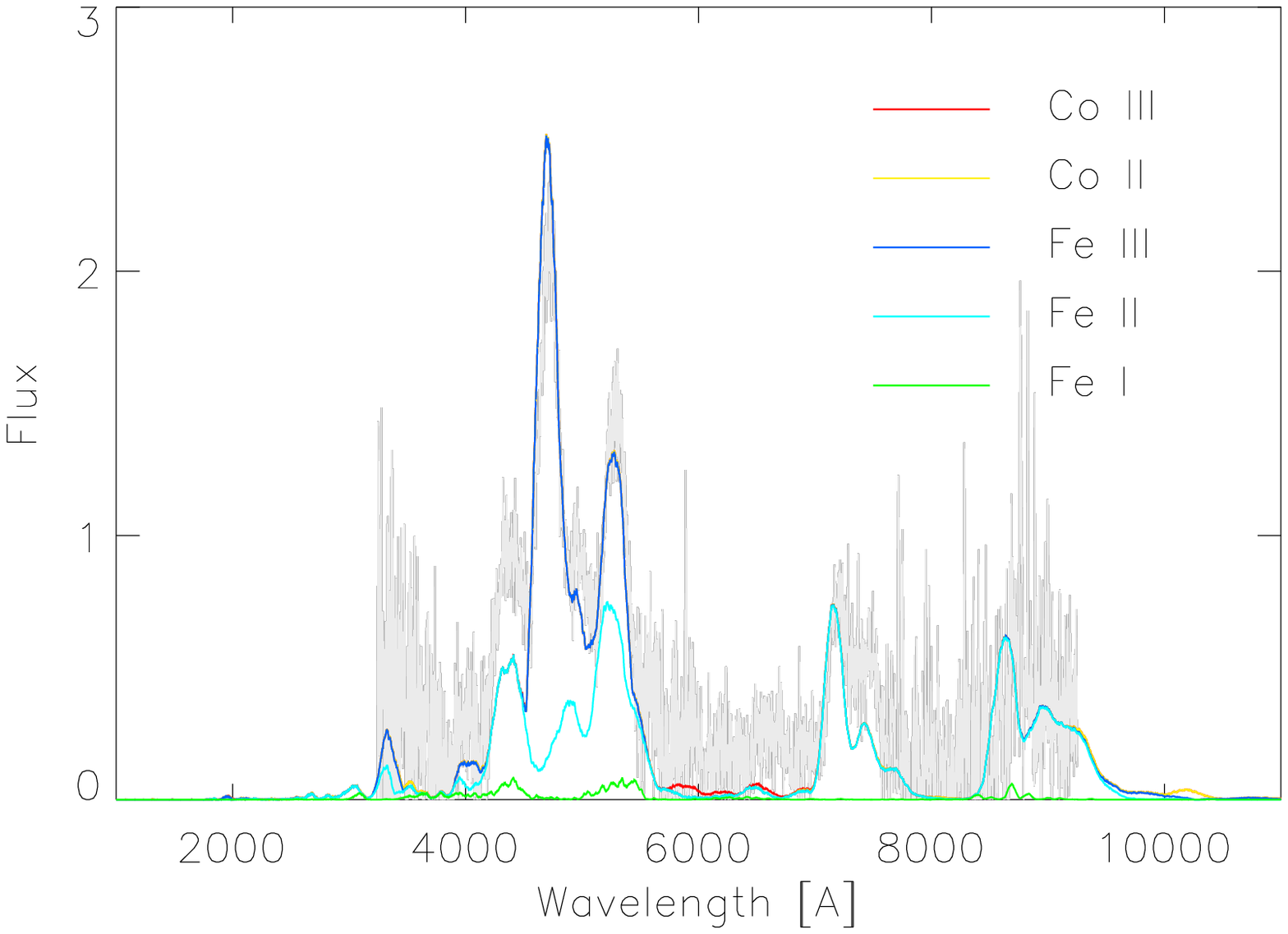}
\end{center}
\caption{The spectrum of SN 2005cf at about 338 days after the
  explosion \citep{Wang09} is shown in black (upper panel) or light
  grey (lower panel). The spectrum was
  scaled by a constant. The coloured curves were
  produced using {\sc nero} on W7. Upper panel: the red
  line shows the total flux. In general the
  agreement is good. Lower panel: the flux of Co {\sc iii}
  (red), Co {\sc ii} (orange), Fe {\sc iii} (dark blue),
  Fe {\sc ii} (light blue) and Fe {\sc i} (green) is shown
  separately. Co features can be observed around 6000 \AA\ and around
  10000 \AA . Their strength may be underestimated in our calculation.}
\label{05cf338}
\end{figure}

To further test the reliability of {\sc nero}, we calculate synthetic
spectra for a W7 model \citep{Nomoto84}, which is expected to reproduce the spectra of
'normal' SNe Ia, although this has never been tested at intermediate epochs. We compare our synthetic spectra
to SN Ia 2005cf \citep[][]{Garavini07,Wang09}, which can be regarded as a proto-type of 
'normal' SNe Ia. The calculations are performed at 47, 94 and 338 days after the
explosion (see Figures \ref{05cf47}, \ref{05cf94} \& \ref{05cf338}). 

At 47 days after the explosion the
agreement between the synthetic and the observed spectrum is
acceptable, given that Co collisional data and forbidden lines are
poorly known. It is interesting to note that the prominent feature at
$\sim$ 8500 \AA , which is usually thought to be a Ca {\sc ii} P-Cygni
profile, could
be strongly influenced by Co {\sc ii} emission in the observed SN Ia
at this epoch. The
collisional data for Co {\sc ii} are very poor and it is not unlikely that we
underestimate (or overestimate) the Co emission in our calculation. Since
important Co collisional data is missing, this remains a speculation.

\begin{figure} 
\begin{center}
\includegraphics[width=8.5cm, clip]{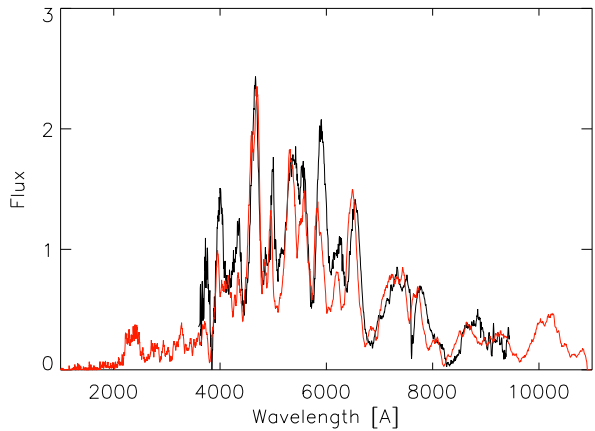}
\end{center}
\caption{The spectrum of SN 2005cf at about 94 days after the
  explosion \citep{Garavini07} is shown in black. The spectrum was
  scaled by a constant. The red curve was produced using RTJ on
  W7. The reproduction of the 4600 \AA\ feature in the RTJ calculation
is much better than in the {\sc nero} calculation. Other features,
for example around 4000 \AA\ and 5900 \AA\ are reproduced better using
{\sc nero}. In general, the agreement with the
observation is excellent.}
\label{cW794anders}
\end{figure}

At 94 days after the explosion there is serious disagreement between
the observed and synthetic {\sc nero} spectrum around 4600 \AA
. Interestingly, in the synthetic spectrum this feature is dominated by Fe {\sc iii} (see Figure
\ref{05cf94}), while another feature, which matches the
observations well, at $\sim$ 5900 \AA\ is dominated by Co
    {\sc iii}. The RTJ spectrum of W7 (see Figure \ref{cW794anders}) fits the 4600 \AA\ much better,
    but under-produces the flux around 5900 \AA . Therefore, there
    seems to be a problem with the ratio of the Fe {\sc iii} and Co
    {\sc iii} line emission in both calculations. Although, one should not expect that the W7 model can
    reproduce the spectra of SN 2005cf in all details, this could mean
    that our atomic data for Fe are inaccurate or that our
    approximations for example for the excited state ionisation
    cross-sections are too simple. In any case, the
    lack of reliable Co {\sc ii} \& {\sc iii} data poses a serious problem for
    all SN Ia spectral calculations between maximum light and $\sim$
    150 days after the explosion before most of the Co has decayed
    into Fe. It has to be hoped
    that these data will be available in
    the near future. 

At 338 days after the explosion the flux is dominated by Fe
emission lines and the agreement of the synthetic and the observed spectrum is
good. At those epochs the nebular spectra of 'normal' SNe Ia are
dominated by three prominent Fe features at roughly 4400, 4700 and
5300 \AA . The Fe 'trident' is shaped by the
ionisation state of the Fe core, which strongly depends on both the
density of the core and the ratio of radioactive and stable iron. Also mixing with
light and intermediate mass elements can influence the relative
abundance of Fe ions.

 While the 4700 \AA\ feature is dominated by [Fe {\sc iii}]
emission, the 4400 \AA\ feature is made from [Fe {\sc i}] and [Fe {\sc
    ii}]. In our synthetic spectrum this feature is underestimated and it
may well be that it contains more contribution from [Fe {\sc i}] in
the observed SNe Ia than predicted in our simulation. It is important to note that very small fractions
of Fe {\sc i} ($\sim$ 0.01\%) are sufficient to cause observable [Fe
  {\sc i}] lines. Such
small fractions of Fe {\sc i} can survive even when Fe {\sc iii} is
present, strongly depending on photo-ionisation, recombination and
possibly on charge
exchange processes. This makes an accurate prediction of [Fe {\sc i}]
features difficult, at least at these epochs. The 5300 \AA\ feature
contains both [Fe {\sc ii}] and [Fe {\sc iii}] and shows also a [Fe {\sc
    i}] contribution. There is almost no Co emission, except weak [Co
{\sc iii}] lines observed around 6000 \AA\ and [Co {\sc ii}] emission
at about 10000 \AA  . Since we have no collision strengths for
Co lines from the literature, their strength may be underestimated in
our simulation. The RTJ and the RTM spectra of W7 at 338 days after
the explosion are similar to the {\sc nero}
spectrum and are shown in the previous section.

\section{Discussion}
\label{disc}
A comparison to the radiation transport code of \citet{Jerkstrand11} has shown 
excellent agreement. This indicates that both codes work properly
within the uncertainties of the atomic data. This is an interesting
result, since they have
been developed completely independent from each other and rely on a different numerical
approach. 

The comparison to the nebular code of
\citet{Mazzali01} has also shown acceptable agreement, especially for pure Fe
cores, which is important for SNe Ia.

For SNe Ic, where several elements like C, O, Na, Mg, Si, S, Ca
and Fe are important for the formation of the nebular spectra, we notice some differences. Most importantly,
photo-ionisation influences the mass estimated for certain elements
like Na and Mg. Also, there is some disagreement regarding the main
properties of the SN core (total and $^{56}$Ni mass). 

It is well known, that mixing or a separation of the elements on
small or large scales in SN ejecta can have strong influence on the resulting
nebular spectra. Since this paper intends to compare codes, we do not 
study this effect in detail. However, it was demonstrated that the
uncertainty caused by the mixing of the ejecta is comparable to the uncertainty
caused by using the different codes, at least for modelling SNe Ib/c. 
A treatment of the mixing
problem in SNe II has been presented by
\citet{Kozma98,Jerkstrand11}. In 'normal' SNe Ia this problem is less
severe, since the core is dominated by $^{56}$Ni decay-products.

By comparing synthetic Ia spectra to observations of
the proto-typical 'normal' SN Ia 2005cf we have shown
that the synthetic spectra produced with {\sc nero} look reasonable and are likely reliable within the
uncertainties of the
atomic data. Most of the observed spectral features have been
identified to result from either
Fe or Co emission. 

At epochs between 50 and 150 days after the
explosion poorly known Co data pose severe problems for
spectral modelling of SNe Ia. Atomic data are essential for calculating SN spectra. Especially for electron
collisions of all kinds the available data are often inaccurate or
incomplete. 

Apart from time-dependent effects, {\sc nero} treats all the radiation
transport effects commonly thought to be important for the formation
of SN spectra in full NLTE. Therefore, {\sc nero} calculations are especially
interesting for intermediate epochs,
since so far SN spectral calculations at 50 $-$ 200 days after the
explosion have been extremely rare. Also, the
nebular phase between roughly 200 and 500 days after the explosion can
be studied.

Possibly, {\sc nero} could be used to calculate (pre-) maximum spectra
by imposing an estimated lower boundary flux at appropriate radii, as it is done in photospheric codes
\citep[e.g.][]{Mazzali93}. With respect to purely photospheric
codes, a treatment with {\sc nero} would include the effect of net
emission above the lower boundary, which could then be set at lower
velocities than in previous approaches. This may increase the accuracy
of the (quasi) photospheric approach considerably. We plan to
investigate this possibility in the near future.

\section{Summary and Conclusion}
\label{sum}
In this paper we presented a new NLTE radiation transport code, which 
can be used to calculate synthetic spectra for all types
of SNe at intermediate and late epochs. Our treatment of intermediate
epochs opens a new window for SN spectral analysis. Currently, {\sc nero} is working
in spherical symmetry, but a three-dimensional version may be
available in the future. In its one-dimensional version the code
can be used for spectral modelling of observed SN spectra or for calculating synthetic spectra
of (approximately) spherically symmetric SN explosion models.

\bibliography{pap7}

\end{document}